\documentclass{PoS}

\title{Uncertainties of QCD predictions
for Higgs boson decay into  bottom quarks
at NNLO and beyond \thanks{Supported by the  RFBR
08-01-00686 and the President of RF NS-1616.2008.2, NS-378.2008.2 Grants.}}

\ShortTitle{Uncertainties of QCD predictions for Higgs boson decay into
$\overline{b}b$  }

\author{\speaker{A~.L~.Kataev } \\
Institute for Nuclear Research of  Russian  Academy of Sciences, 1171312,
Moscow \\
E-mail:   \email{kataev@ms2.inr.ac.ru}}

\author{V.~T.~Kim \\
St. Petersburg  Nuclear Physics Institute of Russian Academy of Sciences
 188300,  Gatchina  \\
  E-mail: \email{ kim@pnpi.spb.ru}}

\abstract{ The importance of detailed studies of 
theoretical QCD predictions
for the decay width of the Standard Model Higgs boson into bottom quarks,
in the case when $\rm{M_H\leq 2M_W}$, is emphasized.
 The effects of higher order
perturbative QCD corrections up to order $\alpha_s^4$-terms are considered.
The resummation of the
 $\pi^2$ terms resulting from analytical continuation, typical of Minkowskian quantities,
is undertaken for the   $\Gamma(H\rightarrow\overline{b}b)$ decay width.
The uncertainties in the calculation of this  decay width
are  analyzed.
}

\FullConference{XII Advanced Computing and Analysis Techniques
in Physics Research\\
		 November 3-7 2008\\
		 Erice, Italy}

\begin{document}

\section{Introduction}

Production cross-sections and decay widths of the Standard Electroweak
Model Higgs boson are nowadays among the most extensively analyzed
theoretical quantities (for recent reviews, see, e.g.,
\cite{Djouadi:2005gi}, \cite{Assamagan:2004mu}).
Indeed, the  main hope
of scientific community is that this essential ingredient of the Standard
Model may be discovered, if not at Fermilab Tevatron, then at the forthcoming
LHC experiments at CERN.
There is great interest in the ``low-mass'' region
114.5 GeV$\leq \rm{ M_H}\leq 2\rm{M_W}$, because a ``low-mass'' Higgs boson
is heavily favored by Standard Model analysis of the available  precision data.
The lower  bound, 114.5 GeV, was obtained  from the direct searches of Higgs boson at the
LEP2 $e^+e^-$-collider primarily through  Higgs boson decay  into a
$\overline{b}b$ pair.
The decay mode $H \rightarrow \overline{b}b$, which is  dominant for the
``low-mass'' region,  is important for Higgs boson searches in certain
associated (semi-inclusive) Higgs boson production processes at the Tevatron and the LHC.
It is also the main decay mode for diffractively produced Higgs boson searches in CMS-TOTEM and,
possibly, FP420 experiments.

It should be stressed, that the uncertainties in
$\Gamma(H\rightarrow \overline{b}b)$,   analytically calculated in QCD
using the  $\overline{\rm{MS}}$-scheme at the
$\alpha_s^4$-level  \cite{Baikov:2005rw},
dominate the
theoretical uncertainty for the branching ratio
of $H\rightarrow \gamma\gamma$ decay, which
is considered to be the most important process in
searches for a ``low  mass'' Higgs boson by CMS and ATLAS collaborations at the LHC.
Moreover, since at present the QCD corrections to the  Higgs boson
production cross-sections at  Tevatron and LHC are known at
the next-to-next-to-leading order (NNLO)
and, partly, even beyond  (see, e.g., \cite{Kidonakis:2005kz}),
it becomes important to understand how to estimate the
theoretical error-bars of  QCD-predictions for both the
production cross-sections and Higgs-boson  decay widths, which are also
calculated in QCD beyond the next-to-leading order (NLO) level.
 Here we will focus on the analysis
of the concrete uncertainties  of the QCD predictions for
$\Gamma_{\rm{H\overline{b}b}}=\Gamma(H\rightarrow \overline{b}b)$, including
those which come
from the on-shell mass  parameterizations of this quantity
(previous related discussions see in
\cite{Kataev:1992fe}-\cite{Kataev:2008ym}) and  from the
resummations of the   $\pi^2$ terms, typical of the
Minkowskian  region (see \cite{Gorishnii:1983cu}- \cite{Bakulev:2008hx}).
\section{QCD expressions for $\Gamma_{\rm{H\overline{b}b}}$ in the
$\overline{\rm{MS}}$-scheme}

\subsection{Basic results in terms of running $b$-quark mass}

The  QCD  prediction  for
$\Gamma_{\rm {H\overline{b}b}}$ in the $\overline{\rm{MS}}$-scheme is of the form
\begin{equation}
\label{MS}
 \Gamma_{\rm{H\bar{b}b}}
  =\Gamma_0^{b}\,
    \frac{{\rm\overline{m}_b^2(M_H)}}
         {\rm{ m_b^2}}\,
     \bigg[1+\sum_{i\geq 1}
              \Delta{\rm \Gamma_i}\,
               a_s^i(\rm{M_{H}})
     \bigg]\,.
\end{equation}
Here
$\Gamma_0^{b}=3\sqrt{2}/{8\pi}\rm {G_F M_{H}{m}_b^2}$,
$\rm\overline{m}_b$ and $\rm{M_{H}}$ are the pole $b$-quark  and Higgs
boson masses,  $a_s(\rm{M_{H}})=\alpha_s(\rm{M_{H}})/\pi$ and
$\rm\overline{m}_b(M_H)$
are the QCD  running parameters, defined
in the $\overline{\rm{MS}}$-scheme.  The coefficients
$\Delta\rm{\Gamma}_i$  can be expressed through   
the 
sum of the following contributions:  the
positive contributions $\rm{d_i^{E}}$, calculated directly
in the Euclidean region, and the ones proportional to $\pi^2$-
factors, which are typical for the Minkowski time-like region.

The corresponding expressions were derived at 
the $\alpha_s^4$-level in Ref.  \cite{Chetyrkin:1997wm}
and have the following form:
\newpage
\begin{eqnarray}
 \Delta{\rm\Gamma_1}
  &=&\
     \rm{d^{\rm{E}}_1}=\frac{17}{3}\,;
 \label{1}\\
 \Delta{\rm\Gamma_2}
  &=&
     \rm{d^{\rm E}_2}-\gamma_0(\beta_0+2\gamma_0)\pi^2/3\,;
 \label{2}\\
 \Delta{\rm\Gamma_3}
  &=&
     \rm{d^{\rm E}_3}-\big[\rm{d^{\rm E}_1}(\beta_0+\gamma_0)
(\beta_0+2\gamma_0)
    +\beta_1\gamma_0+2\gamma_1(\beta_0+2\gamma_0)\big]\pi^2/3\,;
 \label{3}\\
 \Delta{\rm\Gamma_4}
  &=&
     \rm{d^{\rm E}_4}-\big[\rm{d^{\rm E}_2}(\beta_0+\gamma_0)
(3\beta_0+2\gamma_0)
    +\rm{d^{\rm E}_1}\beta_1(5\beta_0+6\gamma_0)/2
    \nonumber \\
  &+&{}
     4\rm{d^{\rm E}_1}\gamma_1(\beta_0+\gamma_0)
    +\beta_2\gamma_0+2\gamma_1(\beta_1+\gamma_1)
    +\gamma_2(3\beta_0+4\gamma_0)\big]\pi^2/3
    \nonumber \\
  &+&{}
     \gamma_0(\beta_0+\gamma_0)(\beta_0+2\gamma_0)
             (3\beta_0+2\gamma_0)\pi^4/30\,,~
  \label{4}
\end{eqnarray}
where the ${\rm n_f}$-dependence of ${\rm d^{\rm E}_i}$ ($2\leq i\leq 4$)
was  evaluated in \cite{Gorishnii:1990zu}, \cite{Chetyrkin:1997wm}
and  \cite{Baikov:2005rw} (the detailed results are presented
in  \cite{Kataev:2008ym}).
The coefficients $\beta_i$ and $\gamma_i$ are the perturbative coefficients
of the QCD renormalization group (RG)  $\beta$-function and mass  anomalous
dimension  function $\gamma_m$ of the ${\rm MS}$-like schemes.
The  QCD $\beta$-function will be  considered at
the 5-loop level:
\begin{equation}
 \label{beta}
  \frac{da_s}{d\ln\mu^2} =
    \beta(a_s) =
    -\beta_0\,a_s^2
    -\beta_1\,a_s^3
    -\beta_2\,a_s^4
    -\beta_3\,a_s^5
    -\beta_4\,a_s^6
    +O(a_s^7)\,.
\end{equation}
The expressions for $\beta_0$ and $\beta_1$ are well-known and are
scheme-independent. The coefficients $\beta_2$ and $\beta_3$ were
analytically evaluated in  \cite{Tarasov:1980au} and
\cite{vanRitbergen:1997va} and confirmed by independent  
calculations at the level of  3-loops \cite{Larin:1993tp} and 4-loops  
\cite{Czakon:2004bu}.  
 The 5-loop coefficient  $\beta_4$ is still unknown, and will be estimated using the
 Pad\'e approximation procedure, developed in  \cite{Ellis:1997sb}.
 The mass anomalous
dimension function is defined as
\begin{equation}
 \label{mass}
  \frac{d{\ln}\overline{\rm m}_b}
       {d{\ln}\mu^2} =
   \gamma_m(a_s) =
       -\gamma_0\,a_s
       -\gamma_1\,a_s^2
       -\gamma_2\,a_s^3
       -\gamma_3\,a_s^4
       -\gamma_4\,a_s^5
       +O(a_s^6)\,.
\end{equation}
 The 4-loop correction  $\gamma_3a_s^4$  was independently
 calculated in \cite{Chetyrkin:1997dh} and
~\cite{Vermaseren:1997fq}.  At  5-loops    $\gamma_4$  may
be modelled using the Pad\'e approximation procedure of 
 Ref.\cite{Ellis:1997sb} mentioned above.
Note that the consideration of the explicit ${\rm n_f}$ dependence  
of the 5-loop coefficients of Eq.(\ref{beta}) and Eq.(\ref{mass}) presented in Ref.\cite{Kataev:2008ym}
strongly suggests 
that the available  Pad\'e estimate  of the  $\gamma_4$-coefficient
contains more uncertainties
than the Pad\'e estimate of the coefficient $\beta_4$,  given in
Ref.\cite{Ellis:1997sb}.
This conclusion is supported  in part
by the fact that  
the Pad\'e estimated value
of the
  ${\rm n_f^3}$-part  of  $\gamma_4$ \cite{Ellis:1997sb}
is over 3 times smaller,  than the
result of the explicit analytical calculation of
Ref. \cite{Ciuchini:1999cv}.
It should be stressed, however, that the uncertainties
of the estimated  5-loop contributions to the QCD $\beta$-function
and mass anomalous dimension function $\gamma_m$ are not so important in the
definition of the running of the $b$-quark mass from the
pole mass $\rm{m_b}$ to the pole mass of Higgs boson $\rm{M_H}$. This effect
of running is described by the
solution of the following RG equation:
\begin{equation}
\label{running} {\rm\overline{m}_b^2(M_H)}
 ={\rm\overline{m}_b^2(m_b)}
  \exp\bigg[-2\int_{a_s(\rm{m_{b}})}^{a_s(\rm{M_H})}
\frac{\gamma_m(x)}{\beta(x)}dx\bigg]=
{\rm \overline{m}_b^2(m_b)}
\bigg (\frac{a_s(\rm{M_{H}})}{a_s(\rm{m_b})}\bigg)^{2\gamma_0/\beta_0}
\bigg(\frac{AD(a_s(\rm{M_ H}))}
{AD(a_s(\rm{m_b}))}\bigg)^2~
\end{equation}
where $AD(a_s)$ is a polynomial  of 4-th order in the QCD expansion
parameter $a_s=\alpha_s/\pi$ (see
\cite{Kataev:2008ym}).
We will complete this section by presenting the numerical values
of Eq.(\ref{1})-Eq.(\ref{4}) in the case of $\rm{n_f=5}$ active quark
flavours  and comment on the significance
of other QED and QCD contributions to
$\Gamma_{\rm{H\overline{b}b}}$, first  considered  in
Refs.\cite{Kataev:1992fe},\cite{Kataev:1997cq},
\cite{Larin:1995sq},
\cite{Chetyrkin:1997vj}.
In the Higgs boson masses region of interest, the expression
for Eq.(\ref{MS}) may be expressed as
\begin{eqnarray}
\label{corr}
\Gamma_{\rm{H\bar{b}b}}
 &=&\Gamma_0^{b}\,
    \frac{{\rm\overline{m}_b^2(M_H)}}
         {\rm{ m_b^2}}\,
     \bigg[1+\sum_{\rm{i}\geq 1}
              \Delta{\rm \Gamma_i}\,
               a_s^{\rm i}(\rm{M_H})
     \bigg] \\ \nonumber
&=&\frac{3\sqrt{2}}{8\pi}{\rm G_F}\rm{ M_H}{\rm\overline{m}_b^2(M_H)}
\bigg[1+ 5.667{\it a_s}(\rm{M_H})+29.15 {\it a_s}(\rm{M_H})^2+
41.76{\it a_s}(\rm{M_H})^3-
825.7{\it a_s}(\rm{M_H})^4\bigg]
\end{eqnarray}
Substituting the value  $a_s(\rm{M_H})\approx 0.0366$ (which corresponds
to $\alpha_s(\rm{M_H}=120~{\rm GeV})\approx 0.115$) into Eq.(\ref{corr}),
and decomposing the coefficients in the Minkowskian series
into  Euclidean  contributions and
Minkowskian-type  $\pi^2$-effects,  one can get from
the work of Ref.\cite{Baikov:2005rw}
the following numbers
\begin{eqnarray}
\label{decomposition}
\Gamma_{\rm{H\bar{b}b}}&=& \Gamma_0^{b}
\frac{\rm{\overline{m}_b^2(M_H)}}{\rm{m_b^2}}\bigg[
1+0.207+0.039+0.0020-0.0015\bigg] \\ \nonumber
&=&\Gamma_0^{b}\frac{\rm{\overline{m}_b^2(M_H)}}{\rm{m_b^2}}
\bigg[1+0.207
+(0.056-0.017)+(0.017 -0.015)+(0.0063-0.0078)~\bigg]
\end{eqnarray}
where the negative numbers  in the round  brackets come from the effects
of analytical continuation. Having a careful look at Eq.
(\ref{decomposition}) we may conclude that in the Euclidean region
the perturbative series is well-behaved and the 
$\pi^2$-contributions typical of the Minkowskian region are also decreasing from order to order.
However, in view of the strong   interplay between these
two effects in the third and fourth terms, the latter
ones are becoming numerically equivalent.
This feature spoils the convergence
of the perturbation series in the Euclidean region. Therefore,
{\bf to improve the precision} of the
 perturbative prediction in the Minkowskian region
it seems natural to {\bf  sum up these $\pi^2$- terms} using
the ideas, developed in the 80s (see, e.g.,  \cite{Yndurain:1980qg}, \cite{Pennington:1981cw},
\cite{Krasnikov:1982fx} and \cite{Radyushkin:1982kg}).
Due to the works discussed, e.g., in  \cite{Shirkov:2000qv},
these ideas now have a more solid theoretical background.
We will describe  some applications
of these resummation procedures to $\Gamma(H\rightarrow \overline{b}b)$
later on. Here we stress
that  the truncated perturbative expansions
of Eq.(\ref{corr}) have some additional uncertainties. These include
$\rm{M_H}$  and   $t$-quark mass dependent QCD
\cite{Larin:1995sq},
\cite{Chetyrkin:1997vj}  and  QED
\cite{Kataev:1997cq} contributions:
\begin{equation}
\Delta \Gamma_{\rm{H\bar{b}b}}=
\frac{3\sqrt{2}}{8\pi}\rm{G_F}M_{H}\overline{m}_b^2(M_H)\bigg[\Delta_{\rm{t}}
+\Delta^{\rm{QED}}\bigg]
\end{equation}
where  $\Delta_{\rm{t}}$ and $\Delta^{\rm{QED}}$
is defined following  Refs. \cite{Chetyrkin:1997vj}, \cite{Kataev:1997cq}
as
\begin{eqnarray}
\Delta_{\rm{t}}&=&
\overline{a}_s^2\bigg((3.111-0.667L_t)+\frac{\rm{\overline{m}_b^2}}{\rm M_H^2}
(-10+4L_t+\frac{4}{3}ln(\rm{\overline{m}_b^2/M_H^2}))\bigg)
\\ \nonumber
&+&\overline{a}_s^3\bigg(50.474-8.167L_t-1.278L_t^2\bigg)
+\overline{a}_s^2\frac{\rm{M_H^2}}{\rm{m_t^2}}\bigg(0.241-0.070L_t\bigg) \\ \nonumber
&+&
X_t\bigg(1-4.913\overline{a}_s+\overline{a}_s^2(-72.117-20.945L_t)\bigg)
\end{eqnarray}
$L_t=ln(\rm{M_H^2/m_t^2})$, $X_t={\rm G_Fm_t^2}/(8\pi^2\sqrt{2})$, $\rm{m_t}$
is the  $t$-quark pole mass, $\rm{\overline{m}_b}=\rm{\overline{m}_b(M_H)}$
\begin{equation}
\Delta^{\rm{QED}}=\bigg(0.472-3.336\frac{\rm{\overline{m}_b^2}}{\rm{M_H^2}}\bigg)a
-1.455a^2+1.301aa_s
\end{equation}
Using $a=\alpha(\rm{M_H})/\pi$=0.0027 ( $\alpha(\rm{M_H})^{-1}\approx129$),
$\rm{m_t}=175~{\rm GeV}$, $\rm{M_H}=120~{\rm GeV}$,
$\rm{\overline{m}_b}=2.8~{\rm GeV}$,
$\rm{G_F}=1.1667\times 10^{-5}~{\rm GeV}^{-2}$ we get
\begin{eqnarray}
\label{H1}
\Delta_{\rm{t}}&=&\bigg[4.84\times10^{-3}-1.7\times10^{-5} \\ \label{H2}
  &+&2.27\times 10^{-3}  +1.85\times 10^{-4} \\ \label{t}
&+&3.2\times 10^{-3}-5.75\times 10^{-4}-2.42 \times10^{-4}\bigg] \\
\label{QED}
 \Delta^{\rm QED}&=&\bigg[ 1.1\times 10^{-3}-4.5\times 10^{-6}-9\times 19^{-6}-
1.2\times 10^{-4}\bigg]
\end{eqnarray}
Comparing the numbers presented in Eq.(\ref{decomposition}) and
Eq.(\ref{H2})-Eq.(\ref{QED}), we conclude that {\bf it seems
more natural} to take into account  order $\alpha_s^4$-terms
in Eq.(\ref{decomposition}) {\bf only after} the {\bf possible discovery} of
the Standard Model Higgs
boson . Indeed, one can see, that even for the light Higgs boson
 the  numerical values of the
order $\alpha_s^4$-contributions to Eq.(\ref{decomposition}) are
comparable with the leading  $\rm{M_H}$- and $\rm{m_t}$- dependent terms in
Eqs. (\ref{H1})-(\ref{t})
and with the leading QED correction in  Eq.(\ref{QED}).
These terms can be neglected at the current level of the experimental precision
of ``Higgs-hunting''.

\subsection{The relations between different definitions of b-quark mass.}
In the discussions above we  used two definitions of the $b$-quark mass,
namely  the pole  mass $\rm{m_b}$ and the running  mass
${\rm\overline{m}_b}$.
At the maximal order we are interested in, i.e.  at  the
next-to-next-to-next-to-leading
order (N$^3$LO),  these two definitions are related in the following way
\begin{equation}
 \label{mb}
 \frac{\rm\overline{m}_b^2(\rm{m_b})}
      {\rm m_b^2}
  = 1-\frac{8}{3}{\it a_s}(\rm{m_b})
     -18.556{\it  a_s}(\rm{m_b})^2
     -175.76{\it  a_s}(\rm{m_b})^3
     -1892 {\it a_s}(\rm{m_b})^4\,
\end{equation}
where the first three coefficients come from the
calculations of Refs. \cite{Chetyrkin:1999qi}, \cite{Melnikov:2000qh}, while
the numerical estimate of the  $\alpha_s^4$-one  is the updated variant of
the estimate of Ref.~\cite{Chetyrkin:1997wm}. It takes
into account the explicit expression for the $O(a_s^3)$ term in Eq.(\ref{mb})
in  the effective charges procedure applied in Ref.\cite{Chetyrkin:1997wm}
and  developed developed previously   in Ref. \cite{Kataev:1995vh}. The
pronounced feature of Eq.(\ref{mb}) is the rapid  growth of the coefficients in
this relation. This property  agrees
with the expectations for the fast increase of
the coefficients of this
perturbation series, revealed in the process of applications  of 
the QCD renormalon
approach  in  Refs.\cite{Beneke:1994sw}, \cite{Bigi:1994em}
(for a discussion  see \cite{Kataev:1994hg}).

There are different points of view concerning the application
of various definitions of $b$-quark mass in phenomenological
studies.
\begin{enumerate}
\item The most popular one  is that in view
of the factorial growth of the coefficients evident in Eq.(\ref{mb}) it is
 better  to avoid application of the pole mass $\rm{m_b}$
and to use instead the $\rm{\overline{MS}}$-scheme
running  $b$-quark mass  $\rm\overline{m}_b(\mu)$ normalized
at the scale $\mu=\rm\overline{m}_b$ (see e.g.
\cite{Beneke:1999fe}, \cite{Kuhn:2007vp}).
\item It is also  possible  to consider  the
invariant $b$-quark mass, which is
related to the running  mass, normalized at the scale
$\mu=\rm{m}_b$   as
\begin{equation}
\label{hatm}
{\rm\hat{m}_b}
 ={\rm\overline{m}_b(\rm{m}_b)}
   \bigg[{\it a_s}(\rm{m_b})^{\frac{\gamma_0}{\beta_0}}
   {AD}(\it{a_s}(\rm{m_b}))
   \bigg]^{-1}\,
\end{equation}
where $AD$ is defined in  Eq.(\ref{running}).
The  concept  of the invariant mass  is rather useful in treating
$\pi^2$-contributions
to   $\Gamma(H\rightarrow b\overline{b})$ (see  \cite{Gorishnii:1983cu}-
\cite{Bakulev:2008hx}).
\item The  pole  $b$-quark mass is frequently  used  in the
MOM-scheme \cite{Jegerlehner:1998zg}.
Within this  prescription  threshold effects  of heavy quarks may
be understood rather easily \cite{Dokshitzer:1993pf}, \cite{Shirkov:1994td},
\cite{Chyla:1995fm}.
Moreover, the concept of
the  pole
$b$-quark mass is commonly applied in considerations  of deep-inelastic
scattering processes \cite{Bierenbaum:2008yu} , and {\bf what is even more
importantly for the LHC}, in {\bf global fits of parton distributions}
\cite{Alekhin:2008hc}. Note, however, that
quite recently the
three-loop transformation
of  the MOM-scheme  to the    ${\rm \overline{MS}}$-scheme
was analysed in Ref.\cite{Chetyrkin:2008jk} where the equivalence
of these two  approaches  outside the threshold region was demonstrated.
 \item Keeping in mind the advantages of both running and pole definitions of the
$b$-quark mass one may analyze the effects of the RG
resummation of $\alpha_s^n{\rm ln}^m(q^2/m_b^2)$-terms  ($m$< $n$)
by comparing theoretical  predictions for concrete  physical quantities,
which depend on ${\rm\overline{m}_b}$ and $\rm{m}_b$.
\end{enumerate}
Note, that for
Eq.(\ref{mb}),  Eq.(\ref{hatm}) and other perturbative
series, discussed  in this work,  we coordinate their  truncated
expressions with the truncation of the inverse logarithmic expressions for
$a_s$. At  the N$^{\rm (k-1)}$LO of perturbation theory
$(1\leq {\rm k}\leq 5)$ they are defined as
\begin{eqnarray}
\label{NLO}
  a_s(\mu^2)_{\rm{ LO}}
  &=&
   \frac{1}{\beta_0{\rm Log_{1}}}~~ a_s(\mu^2)_{\rm{NLO}}
= \frac{1}{\beta_0{\rm Log_{2}}}
 \bigg[1-\frac{\beta_1{\ln({\rm Log_{2}})}}{\beta_0^2{\rm Log_{2}^2}}\bigg] \\
\label{N2LO}
a_s(\mu^2)_{\rm{N^2{LO}}}
   &=&
    a_s(\mu^2)_{\rm{NLO}}
    +\Delta a_s(\mu^2)_{\rm{{N}^2{LO}}}~~~
  a_s(\mu^2)_{\rm{{N}^3{LO}}}
   =
    a_s(\mu^2)_{\rm{{N}^2{LO}}}
    +\Delta a_s(\mu^2)_{\rm{{N}^3{LO}}} \\
\label{N3LO}
a_s(\mu^2)_{\rm{{N}^4{LO}}}
  &=&
    a_s(\mu^2)_{\rm{{N}^3{LO}}}
    +\Delta a_s(\mu^2)_{\rm{{N}^4{LO}}}
 \end{eqnarray}
where  ${\rm Log_{ k}}={\rm ln(\mu^2/\rm{\Lambda_{k}^2}})$,
$\rm{\Lambda_{k}}$ are  the values of the QCD scale  parameter
$\rm{\Lambda^{(n_f)}_{\overline{MS}}}$, extracted from
the experimental data for concrete  physical quantities
taking into account  N$^{\rm (k-1)}$LO  perturbative
QCD corrections, which depend on the number of active quark flavours, $\rm{n_f}$. 
The definitions of  $\Delta a_s(\mu^2)_{\rm{N^{(k-1)}{LO}}}$
$3\leq \rm {k} \leq 5$ in Eq.(\ref{N2LO}) and Eq.(\ref{N3LO})
contain
high-order scheme-dependent coefficients $\beta_{k-1}$
of the QCD $\beta$-function of Eq.(\ref{beta}) and
$\rm{\Lambda^{(n_f)}_{\overline{MS}}}$ as well
(for details see Ref. \cite{Kataev:2008ym}).

\subsection{Explicit expression for $\Gamma_{\rm{H\overline{b}b}}$
in terms of  pole mass}
Consider the  ${\rm\overline{MS}}$-scheme perturbative series  for
$\Gamma_{\rm{H\overline{b}b}}$ from Eq.(\ref{MS}).
To transform  it to the case when instead of the running mass
${\rm\overline{m}_b(M_H)}$   the pole mass $\rm{m_b}$ is  used,
one should make the following steps:
\begin{itemize}
\item  express   ${\rm\overline{m}_b(M_H)}$ in terms of
${\rm\overline{m}_b(m_b)}$ and $\alpha_s(\rm{m}_b)$ by solving
the RG equation  for the running mass, defined
in Eq.(\ref{running});
\item  apply  Eq.(\ref{mb}), which relates the  square of the  running mass
 ${\rm\overline{m}_b(m_b)}$ to the square of the pole mass $\rm{m}_b$
via a perturbative  expansion  in powers
$a_s(\rm{m}_b)=\alpha_s(\rm{m}_b)/\pi$;
\item
reexpress powers of ${\it a_s}(\rm{m}_b)$,
which appear at the first and
and second steps, in terms of powers of ${\it a_s}(\rm{M_{H}})$
using the RG  equation of Eq.(\ref{beta})  for the
QCD coupling constant.
\end{itemize}
The explicit
solutions of  equations  mentioned
above, namely the solutions of
Eq.(\ref{running}) and Eq.(\ref{beta}), were written down  in
Ref.\cite{Kniehl:1994dz} and extended to higher order level
in   Ref.\cite{Chetyrkin:1997wm}. Their application
result in the appearance  in the expressions for
$\Gamma_{\rm{H\overline{b}b}}$
\begin{equation}
\label{OS}
 \Gamma_{\rm{H\overline{b}b}}
 =\Gamma_0^{b}
   \bigg[1
    + \Delta{\rm \Gamma^{b}_1} {\it a_s}(\rm{ M_H})
    + \Delta{\rm \Gamma^{b}_2} {\it a_s}(\rm{ M_H})^2
    + \Delta{\rm \Gamma^{b}_3} {\it a_s}(\rm{M_H})^3
    + \Delta{\rm \Gamma^{b}_4} {\it a_s}(\rm{M_H})^4
    \bigg]~~
\end{equation}
of the RG controllable
$L= {\rm ln(M_H^2/m_b^2)}$-terms , which enter
into the  coefficients   $\Delta{\rm\Gamma^{b}_i}$ in the following
way
\begin{eqnarray}
 \label{G1}
  \Delta{\rm \Gamma_1^{b}}
   &=& 3-2\,L\,; \\
 \label{G2}
  \Delta{\Gamma_2^{b}}
   &=&-4.5202
              -18.139\,L
              +0.08333\,L^2\,; \\
 \label{G3}
  \Delta{\Gamma_3^{b}}
   &=& -316.88
              -133.42\,L
              -1.1551\,L^2
              +0.0509\,L^3\,; \\
 \label{G4}
  \Delta{\Gamma_4^{b}}
   &=& -4366.2
              -1094.6\,L
              -55.867\,L^2
              -1.8065\,L^3
              +0.0477\,L^4~~.
\end{eqnarray}
The numerical values of the calculated logarithmic terms
are not small. However, in the case of
${\rm i\geq 2}$  they tend to  cancel each other in
the final results  for $\Delta{\rm \Gamma_i^{b}}$.

The related variant of  Eq.(\ref{MS}), where the RG-controllable terms
are summed up, may be written down as
\begin{eqnarray} \label{RG}
 \Gamma_{\rm{H\overline{b}b}}
 &=&
 \Gamma_0^{b}
  \bigg(\frac{a_s(\rm{M_H})}
             {a_s(\rm m_b)}
  \bigg)^{(24/23)}
   \frac{AD(a_s(\rm{M_H}))^2}
        {AD(a_s(\rm{m_b}))^2} \bigg[1+\sum_{{\rm i}\geq 1}
              \Delta{\rm \Gamma_i}\,
               a_s^{\rm i}(\rm M_H)
     \bigg]
  \\ \nonumber
  &\times&
  \big(1-\frac{8}{3}{\it a_s}(\rm {m_b})
        -18.556\,{\it a_s}(\rm {m_b})^2
        -175.76\,{\it a_s}(\rm{m_b})^3
        -1892\,{\it a_s}(\rm{m_b})^4
  \big)\,,
\end{eqnarray}
where
\begin{equation}
 AD(a_s)^2=1+2.351\,a_s+4.383\,a_s^2+3.873\,a_s^3-15.15\,a_s^4
\end{equation}
We will   factorize the term   $\rm\Gamma_0^{b}$  out of the expressions
for   Eq.(\ref{OS})
and Eq.(\ref{RG}) as well. These  representations are
rather convenient   for comparing
different parameterizations of
 $\Gamma_{\rm{H\overline{b}b}}$ and of the ratio
$\rm{R(M_H)}= \Gamma_{\rm{H\overline{b}b}}/\Gamma_0^{b}$.

\subsection{Values of the QCD parameters used}
To analyze   the behavior of the
truncated series in  Eq.(\ref{OS}) and Eq. (\ref{RG}) and of  the various
approximations  for the $\rm{R}$-ratio it is necessary to fix definite
values of the QCD parameters
 $\rm{m_b}$,  $\Lambda^{(\rm n_f=4)}_{\overline{\rm MS}}$  and
$\Lambda^{(\rm n_f=5)}_{\overline{\rm MS}}$.
This is done in Table 1.
\begin{table}[h]
\centerline{ \begin{tabular}{|c|p{20mm}|p{20mm}|p{20mm}|} \hline
order  & ${\rm m_b}~{\rm GeV}$ & $\Lambda^{(\rm
n_f=4)}_{\overline{\rm MS}}$~${\rm MeV}$ &
 $\Lambda^{(\rm n_f=5)}_{\overline{\rm MS}}$~${\rm MeV}$ \\
\hline
LO & 4.74 & 220 & 168 \\
NLO & 4.86 & 347 & 254 \\
${\rm N^2LO}$  & 5.02     & 331    & 242 \\
${\rm N^3LO}$  & 5.23     & 333    & 243 \\
${\rm N^4LO}$  & 5.45     & 333    & 241 \\
\hline \end{tabular}}
\caption{The values of the QCD  parameters used.}
\label{Tab}
\end{table}

The LO, N$^{\rm k}$LO ($1\leq \rm{k}\leq 3$) results for  the pole mass
${\rm m_b}$
are taken from Ref.
\cite{Penin:2002zv}, where they were extracted using the relation between the
mass of the $\Upsilon\rm{(1S)}$- resonance, ${\rm m_b}$  and
the ground state
energy  of $\Upsilon\rm{(1S)}$-system. The N$^4$LO estimate  of
${\rm m_b}$ is
our theoretical guess.   The LO, $\rm{NLO}$, $\rm{N^2LO}$
 values   for $\Lambda^{(\rm n_f=4)}_{\overline{\rm MS}}$, given
in Table 1,
come from the
recent {\bf  parton distribution fits} of Ref.~\cite{Martin:2007bv}. At
the  NLO and $\rm{N^2LO}$ level   these numbers  agree with
the  values of
$\Lambda^{(\rm n_f=4)}_{\overline{\rm MS}}$, which
were  extracted from the first N$^3$LO QCD  analysis
of the experimental data, performed in Ref.\cite{Kataev:2001kk}.
The fitted data were the Tevatron experimental data points
for the
$xF_3$ structure function
of the $\nu N$ deep-inelastic
scattering (DIS)  process and were    obtained  by
CCFR collaboration \cite{Seligman:1997mc}.
The      $\rm{N^3LO}$ value of
$\Lambda^{(\rm n_f=4)}_{\overline{\rm MS}}$ in Table 1
is one of the results of Ref.\cite{Kataev:2001kk}.
In view of the indications for a convergence of the fits performed at $\rm{N^4LO}$ revealed in
Ref. \cite{Kataev:2001kk} 
we will use  the same
value of
  $\Lambda^{(\rm n_f=4)}_{\overline{\rm MS}}$ as at the $\rm{N^3LO}$ level.
To get the results for  $\Lambda^{(\rm n_f=5)}_{\overline{\rm MS}}$, given
in the last column of Table 1,
we apply the
 NLO and N$^{2}$LO   matching conditions of
Ref.\cite{Bernreuther:1981sg} (the latter ones were
corrected a bit in Ref.\cite{Larin:1994va}). At the     N$^3$LO
we use  the expressions  from   Ref.~\cite{Kniehl:2006bg}. At
the  N$^{4}$LO  the   analytical relation from
Ref.  \cite{Kniehl:2006bg} was  applied.
Note, that the calculations of Ref. \cite{Kniehl:2006bg}
confirmed the validity of
analogous  expressions,
obtained (partly numerically)
 in
Ref.\cite{Schroder:2005hy}
and  Ref.\cite{Chetyrkin:2005ia}.
The related NLO-N$^4$LO
results for $\alpha_s(\rm M_Z)$ are contained  in the interval (0.118-0.119).
Thus they agree  with the world average value of $\alpha_s(\rm M_Z)$
(for  a recent review see Ref.\cite{Bethke:2006ac}).

\subsection{The comparisons of the renormalization group improved
and the  truncated  pole-mass parameterizations}
\begin{figure}[h]
{\includegraphics[width=56mm]{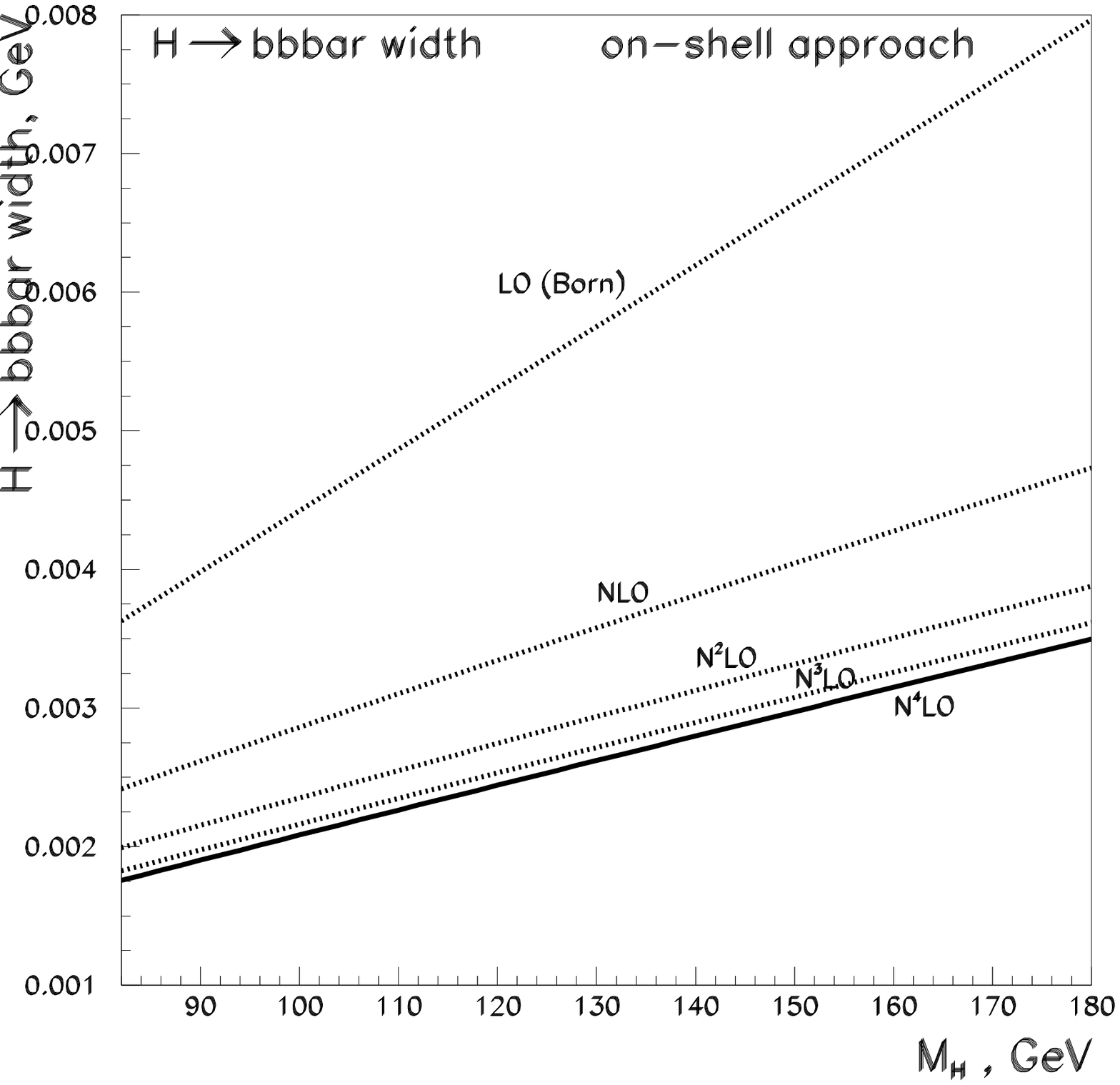}
 \includegraphics[width=55mm]{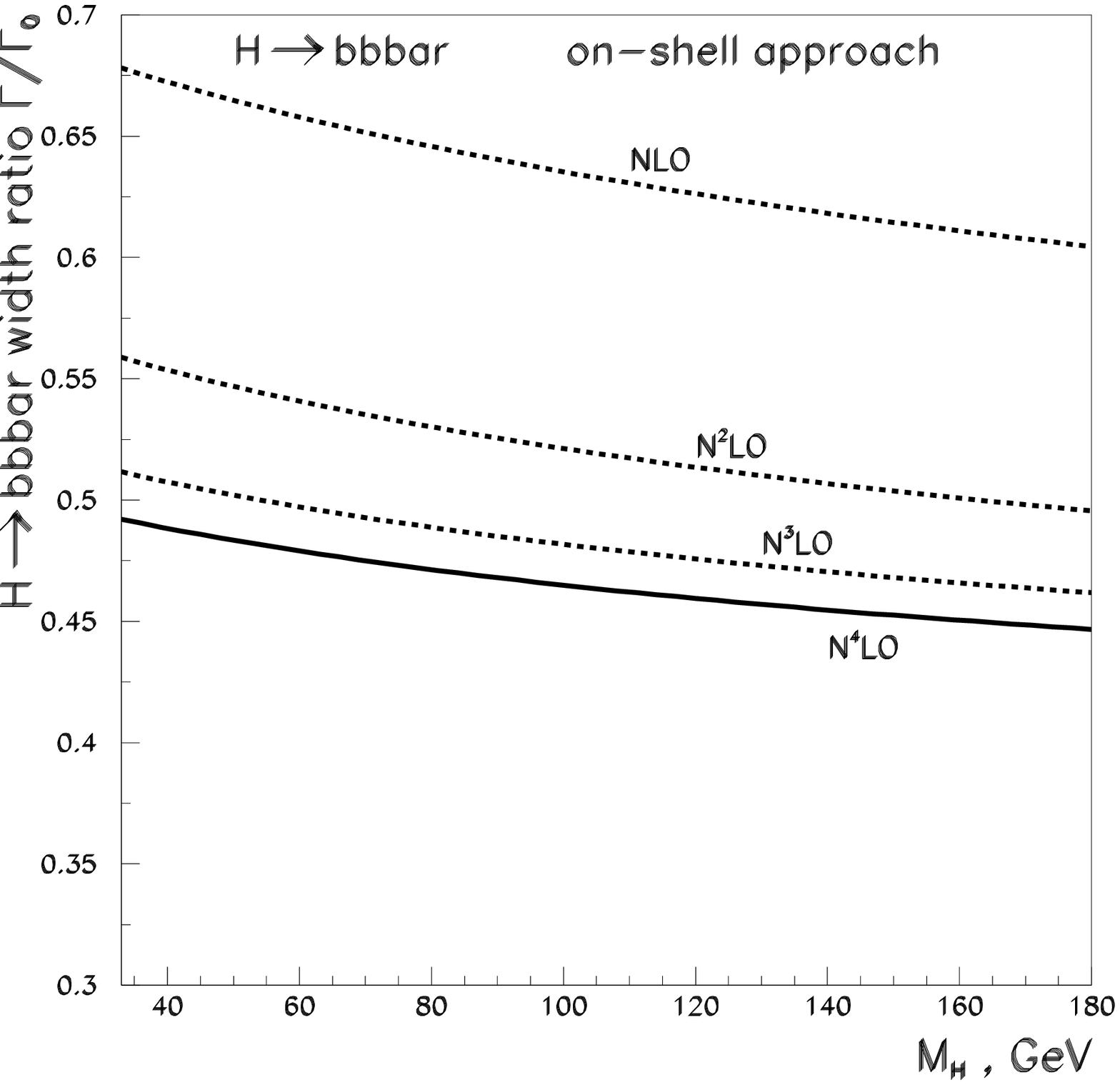}}
 \caption{The quantities analyzed in the pole (or on-shell) mass approach.
 \label{fig1}}
{\includegraphics[width=55mm]{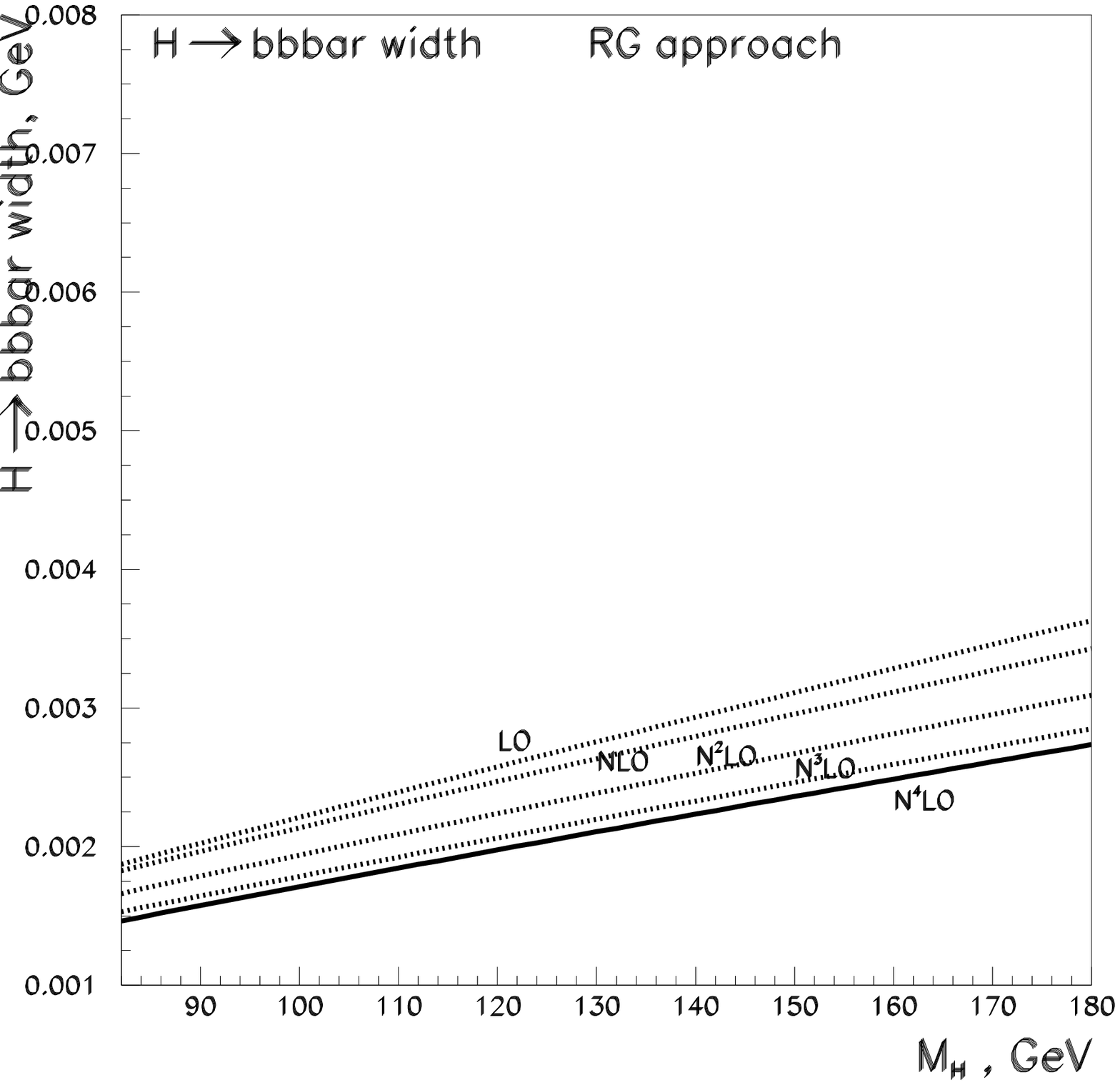}
 \includegraphics[width=55mm]{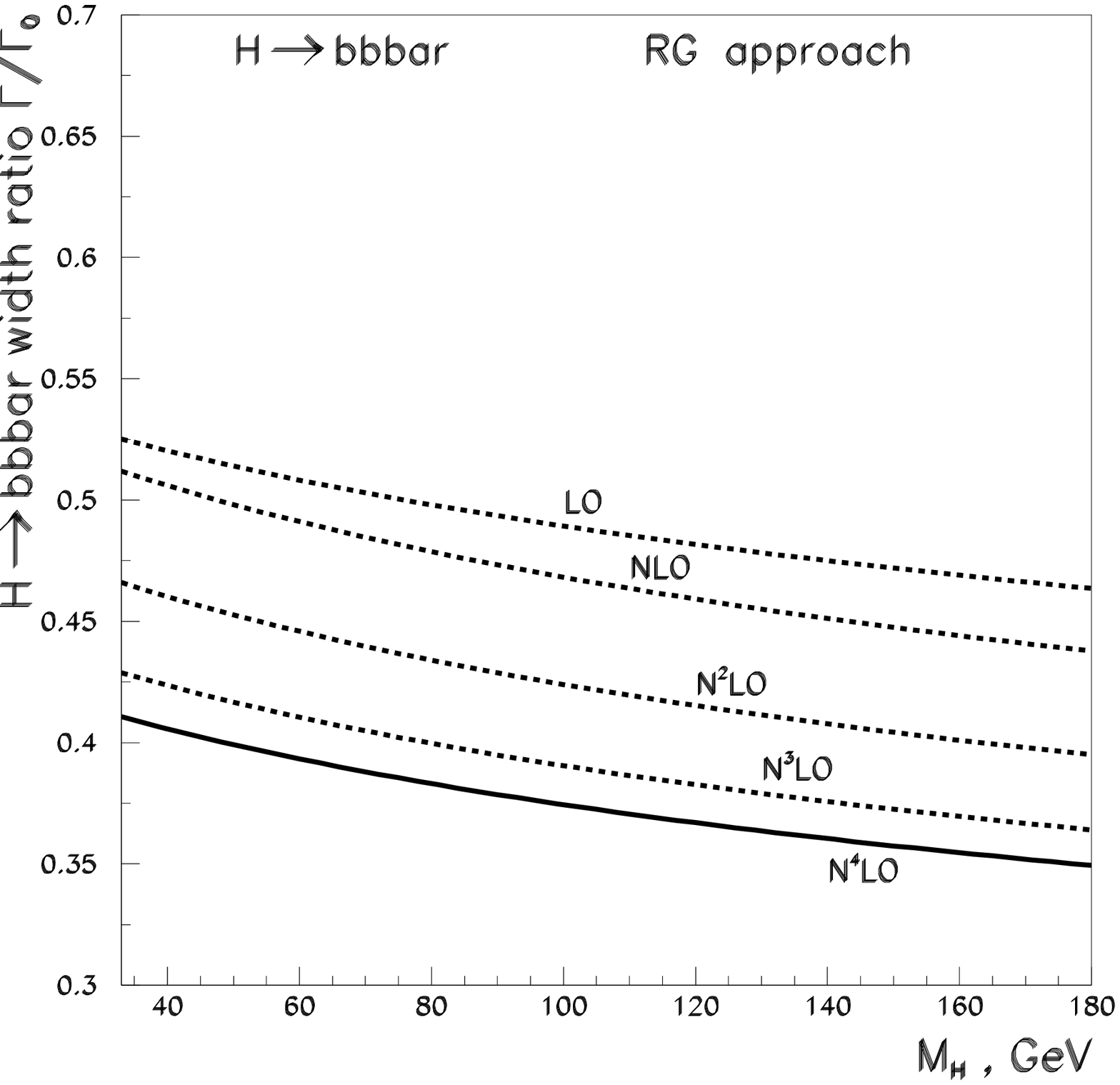}}
  \caption{The quantities analyzed in the approach with explicit
RG-resummation
  \label{fig2}}
\end{figure}

Different perturbative
approximations for two parameterizations of $\rm\Gamma_{H\overline{b}b}$
(see  Eq.(\ref{OS}) and Eq.(\ref{RG})) and of the related
 ratios $\rm{R(M_H)}$
are compared in the plots of Figure 1
and Figure 2.
Looking carefully at  these plots, we conclude that
\begin{enumerate}
\item the partial cancellation of
the
RG-controllable  large
contributions to
Eqs.(2.25)-(2.27),  which are proportional to
$L= {\rm ln(M_H^2/m_b^2)}$ and which appear in  the coefficients of the
pole-mass parameterization of
$\rm\Gamma_{H\overline{b}b}$, leads to a reduction
in the size of perturbative corrections.
Taking them into  account results in a decrease of the difference bewteen 
the behaviour of   the curves of
{\bf Fig.1}
and  {\bf Fig.2}. This feature demonstrates the numerical
importance  of
step-by-step
application of the  RG-resummation
approach.
\item
The cancellations between the contributions proportional to $L$ 
in the coefficient  $\Delta{\Gamma_4^{b}}$ result in the small value of
the   whole $\alpha_s(\rm{M_H})^4$ correction to  the perturbative
expression for
$\rm\Gamma_{H\overline{b}b}$.
This fact demonstrates the    convergence of
different theoretical  approximants.
From a {\bf phenomenological } point of view
this means that at present the $\alpha_s^4$
corrections {\bf may  be  neglected} in the related
computer codes  for calculating some  branching ratios, e.g., for 
$H\rightarrow\gamma\gamma$,
$H\rightarrow\overline{b}b$ and  $H\rightarrow \tau^+\tau^-$ processes.
This conclusion is consistent with our similar  statement, made  in the case of
using running $b$-quark $\rm\overline{m}_b$, but  motivated by
 different theoretical arguments (see the end of {\bf Sec. 2.1}).
\item
The behavior of the RG-resummed  expressions
for $\rm\Gamma_{H\overline{b}b}$ and  $\rm{R_{H\overline{b}b}}$
are more  stable than  in  the case, when RG  summation of
the mass-dependent terms
is not used (see {\bf Fig.1}).
This feature supports the application of the
complete RG-improved parameterization of Eq.(\ref{RG}), which is
closely related  to the one, defined through  the  running $b$-quark mass
(see  Eq.(\ref{corr})).
\item
We observed the existence of a difference
$\Delta\Gamma_{\rm H\overline{b}b}$ between
the truncated pole-mass approach  and the RG-improved
parameterization of
 $\rm\Gamma_{H\overline{b}b}$.
A pleasant feature is that this difference
becomes smaller and smaller in each successive order of
perturbation theory considered.
Indeed, for the phenomenologically interesting value
of Higgs boson mass ${\rm M_H}=120~ {\rm GeV}$ we find that
at the  $\alpha_s^2$-level
$\Delta\Gamma_{\rm H\overline{b}b}\approx 0.7~{\rm MeV}$,
while for the  $\alpha_s^3$-curves it  becomes smaller, namely
$\Delta\Gamma_{\rm H\overline{b}b}\approx 0.3~{\rm MeV}$.
At the $\alpha_s^3$-level of the RG-improved $\rm\overline{MS}$-scheme
series
one has  $\rm\Gamma_{H\overline{b}b}\approx 1.85 ~{\rm MeV}$
for ${\rm M_H}=120~{\rm GeV}$.
For  this scale the value
of  $\rm\Gamma_{H\overline{b}b}$   with the  explicit dependence from
the pole-mass is {\bf $16~\%$} higher, than its RG-improved estimate.
\end{enumerate}
We hope, that further studies may clarify whether it is possible to
formulate more well-defined   procedures for determining the theoretical
errors for this important characteristic of the Higgs boson and for the
predictions of other perturbative QCD series as well. Indeed, even
the $\rm\overline{MS}$-scheme   expression
for $\rm\Gamma_{H\overline{b}b}$ at  the $\alpha_s^3$-level reveal
additional  theoretical uncertainties, which are  related not only to the
the estimates of  theoretical errors  for  the
$\rm\overline{MS}$-scheme parameters
$\rm\overline{m}_b(m_b)$ and $\overline\alpha_s(M_Z)$
(for a discussion see e.g. Ref.\cite{Amsler:2008zzb}).
These not previously  specified   uncertainties result from application of
different approaches  to  the treatment  of  the typical Minkowskian 
$\pi^2$-contributions
in the perturbative expressions for physical quantities. In the case
of $\rm\Gamma_{H\overline{b}b}$ the results from application
of these different procedures  will be compared  below.

\section{Resummations of $\pi^2$-terms}
\subsection{The definitions of the resummed approximants}
As was demonstrated in
Ref.\cite{Baikov:2005rw}, the kinematic  $\pi^2$-terms in the
coefficients of perturbative series  for $\rm\Gamma_{H\overline{b}b}$
become comparable
with the Euclidean contributions starting from
the N$^3$LO $\alpha_s^3$ corrections (see Eq.(\ref{decomposition})).
In general,  $\pi^2$-terms are resulting from
analytical continuation to the Minsowskian  region  of
theoretical expressions for  physical
quantities,   defined through the
the two-point functions of quark or  gluon currents,
which are calculated in the Euclidean region.
These terms are
starting to manifest themselves  from  the N$^2$LO.
In Ref. \cite{Gorishnii:1983cu}
the idea of resummation of
kinematic $\pi^2$-effects of
 Ref.\cite{Krasnikov:1982fx}
and Ref. \cite{Radyushkin:1982kg} was
generalized to the case when the related RG-equation
has the non-zero anomalous dimension. In that work the
case of the N$^2$LO approximation for $\rm\Gamma_{H\overline{b}b}$
was analyzed.
However, since the negative $\pi^2$-contributions to the N$^2$LO correction
to $\rm\Gamma_{H\overline{b}b}$ turn out to be  smaller than the value of
corresponding Euclidean term
(see Eq.(\ref{decomposition})),
the possible development  of the $\pi^2$ resummation procedure was
overlooked by the authors of the works of Ref.  \cite{Gorishnii:1983cu} and
Ref. \cite{Gorishnii:1990zu}, aimed at the study of scheme-dependence
of  $\alpha_s^2$
approximations for this Higgs boson  characteristic.
The appearance of Contour Improved Perturbation
Theory (CIPT)  and its application in the  semi-hadronic decay channel of
the $\tau$-lepton \cite{Pivovarov:1991rh}, \cite{Le Diberder:1992te}
pushed ahead the real interest in the development of
$\pi^2$-resummation  approaches both in theoretical and phenomenological
investigations.

In order to resum the kinematic  $\pi^2$-contributions
to  perturbative   predictions for
$\rm\Gamma_{H\overline{b}b}$  the authors of
Ref. \cite{Broadhurst:1994se}
supplemented
 the CIPT method with the
procedure of
``Naive Non-Abelianization'' (NNA) \cite{Broadhurst:1994se},  commonly
used in the renormalon calculus approach
( a  detailed discussion of this method can be found
in
Ref.\cite{Beneke:1998ui}).
As the result, the
following approximation  for  $\rm\Gamma_{H\overline{b}b}$ was
obtained \cite{Broadhurst:2000yc}:
\begin{equation}
\label{BKM}
\rm\Gamma_{H\overline{b}b}^{\rm{BKM}}
=\rm\Gamma_0^{b}
    \frac{\rm\hat{m}_b^2}
         {\rm m_b^2}\bigg[(\it{a_s}(\rm{M_H}))^{\rm{\nu_0}}{\rm A_0}+
\sum_{n\geq1} (\it{a_s}(\rm{M_H}))^{\rm{\nu_0}}{ \rm{d^{\rm{E}}_n}
 A_n(\it{a_s}(\rm{M_H}))}
    \bigg]
\end{equation}
where the ${\rm A_n}$-functions  are defined as:
\begin{equation}
{\rm A_n}={\rm \frac{1}{\beta_0\delta_n\pi}\big[1+\beta_0^2\pi^2
{\it a_s}^2\big]^{-\delta_n/2} (\it{a_s})^{\rm{n-1}}{\sin}
\big(\delta_{\rm n}{\arctan}(\beta_0\pi {\it a_s})\big)}\,,
\end{equation}
Here
$\rm{\delta_n=n+\nu_0-1}$ and  $\rm{\nu_0}=2\gamma_0/\beta_0$ depends
on the first coefficient of the QCD $\beta$-function
$\beta_0=(11-2/3{\rm n_f})/4$, introduced in Eq.(\ref{beta})
and $\it{a_s}(\rm{M_H})=1/(\beta_0\rm{ ln(M_H^2/\Lambda^2)})$
It should be stressed that the NNA  approach
is dealing
with the leading terms in expansions of perturbative coefficients
for Euclidean quantities in powers of number of flavor ${\rm n_f}$
and it provides the basis of  ``large $\beta_0$-approximation''.
Within this approximation it is assumed that the terms, proportional to
$\beta_0^{\rm p}$ ( where $\rm{p}$  is ``large'', namely
$1\leq \rm{p} \leq \infty$) give a qualitatively good  approximation
for the structure  of the Euclidean  perturbative contributions
to physical quantities under study. Indeed, as was shown
in Ref. \cite{Broadhurst:2002bi} this approach gives  correct
both for  sign  
for order of magnitude estimates of the
perturbative coefficients $\rm{d_i^{E}}$  to various physical
quantities, including ${\rm d_4^{E}}$- contribution to
$\rm\Gamma_{H\overline{b}b}$, defined in Eq.(\ref{4}).
The explicit calculations of
Ref.\cite{Baikov:2005rw} demonstrate, that  at  $\rm{n_f}$=5 the real value
of ${\rm d_4^{E}}$-coefficient is higher, than its   NNA estimate
from  Ref.\cite{Broadhurst:2002bi}  by the factor 5 approximately.
\footnote{Unfortunately, NNA results from  Table I 
of the important work of  Ref. \cite{Baikov:2005rw}  contain misprints.}.

Within this ``large $\beta_0$-approximation''
is seems more consistent to approximate
the perturbative QCD  expansion parameter
$a_s(\rm{M_H})=\alpha_s(\rm{M_H})/\pi$
by  its LO-expression of Eq.(2.20).  Fixing now ${\rm n=0}$ in Eq.(3.2)
 and
expanding ${\rm A_0}$  to the first order in $a_s$,
the authors of Ref.\cite{Broadhurst:2000yc}
got
\begin{equation}
{\rm A_0}= \frac{\rm 1}{\rm b_0{L_{M_H}}^{b_0}}
     \frac{\rm{\sin(b_0~\arctan}({\it\pi}/{\rm L_{M_H}}))}
          {(1+{\it \pi}^2/\rm{L_{M_H}}^2)^{b_0/2}}\,,
\end{equation}
where ${\rm b_0=\nu_0-1}$, ${\rm L_{M_H}=ln(M_H^2/\Lambda^2)}$.
This concrete expression   was first derived in
Ref.~\cite{Gorishnii:1983cu}.

Other ways of resumming $\pi^2$-contributions to $\rm\Gamma_{H\overline{b}b}$
 were considered
within  the framework of Fractional Analytical Perturbation Theory (FAPT)
\cite{Bakulev:2006ex} and its variant  (for a review of
FAPT   see Ref.  \cite{Bakulev:2008td} where
   ``flavor-corrected global FAPT''    was also proposed).
It should be stressed that  the cornerstone of FAPT is the  Analytical
Perturbation Theory approach, which  was developed in the  studies,
initiated  by the work  \cite{Shirkov:1997wi}.

Applying the
 ``large $\beta_0$-expansion'', which is equivalent to the
 choice of the LO expression
for $a_s$, one can get the FAPT analog of Eq.(\ref{BKM}).
It can be written down as
\begin{equation}
\label{1-Fapt}
\rm\Gamma_{H\overline{b}b}^{\rm(1;FAPT)}=
\rm\Gamma_0^{b}
    \frac{\rm\hat{m}_b^2}
         {\rm m_b^2}
      \left[{\mathfrak A}_{\rm\nu_{0}}^{(1)}(\rm{M_H})
          + \sum_{n\geq1}^{4}\rm{d^{\rm{E}}_n}
             \frac{{\mathfrak A}_{n+\nu_{0}}^{(1)}(\rm{M_H})}
                  {\pi^{n}}
      \right]\,.
\end{equation}
{\bf This approximant almost coincides with
the expression of Eq.(\ref{BKM})}. This conclusion was made
in Ref. \cite{Bakulev:2006ex} taking into account 
the following definitions of the functions in Eq.(3.4), namely
\begin{eqnarray}
 {\mathfrak A}_{\rm\nu}^{(1)}(\rm{M_H})&=&\frac{\rm{sin[(\nu-1)arccos(L_{M_H}
/\sqrt{\pi^2+L^2_{M_H}})]}}{\rm{\pi(\nu-1)(\pi^2+L^2_{M_H})^{\nu-1/2}}} \\
{\mathfrak A}_{n+\nu}^{(1)}(\rm{M_H})&=&\frac{\Gamma(\nu)}{\Gamma(n+\nu)}
\bigg(-\frac{\rm{d}}{\rm{d~L_{M_H}}}\bigg){\mathfrak A}_{\rm\nu}^{(1)}(\rm{M_H})
\end{eqnarray}

The FAPT N$^3$LO expression of Ref.(3.4) was given   in
Refs.\cite{Bakulev:2006ex}, \cite{Bakulev:2008td}. It has the following form:
\begin{equation}
\label{3-FAPT}
\rm\Gamma_{H\overline{b}b}^{(\rm{3;FAPT})}
=  \rm\Gamma_0^{b}
    \frac{\rm\hat{m}_b^2}
         {\rm m_b^2}
      \left[{\mathfrak B}_{\nu_{0}}^{(3)}(M_H)
          + \sum_{n\geq1}^{3} \rm{d^{\rm{E}}_n}
             \frac{{\mathfrak B}_{n+\nu_{0}}^{(3)}(M_H)}
                   {\pi^{n}}
      \right]\
\end{equation}
Within the framework of FAPT  the functions
${\rm{\mathfrak B}_{n+\nu_{0}}^{(3)}(M_H)}$ ($0\leq {\rm n} \leq 3$)
absorb the evolution of the $\rm\overline{MS}$-scheme running b-quark mass and
$\pi^2$-contributions
in Eq.(\ref{MS}), which are
proportional to high-order  coefficients of
RG-functions   $\beta(a_s)$ and $\gamma_m(a_s)$
in Eqs.(2.1)-(2.5).

In the ``flavour-corrected (f-c) global FAPT'', proposed in 
Ref.\cite{Bakulev:2008td}, the  FAPT expression
for $\rm\Gamma_{H\overline{b}b}$ is expanded into the functions
${\rm {\mathfrak B}_{l;\rm{d_n}}^{(l)}(s)}$,
which absorb all $\rm{n_f}$ dependence
from  the Euclidean coefficients  $\rm{d^{\rm{E}}_n}$
in Eqs.(2.3)-(2.5).The explicit expression for this approximant
reads  \cite{Bakulev:2008hx}
\begin{equation}
\label{f-c,FAPT}
\rm\Gamma_{H\overline{b}b}^{(\rm{3;FAPT,f-c})}=
\rm\Gamma_0^{b}
    \frac{\rm \hat{m}_b^2}
         {\rm m_b^2}
       \left[{\mathfrak B}_{\nu_{0}}^{(3)}(M_H)
          + \sum_{n\geq1}^{3}
             \frac{{\mathfrak B}_{n+\nu_{0};\rm{d^{\rm{E}}_n}}^{(3)}(M_H)}
                  {\pi^{n}}
     \right]\,
\end{equation}
The more detailed study of the characteristic feature
of  formalizm  \cite{Bakulev:2008td}, \cite{Bakulev:2008hx}
is now in progress  \cite{Bakulev:2009}.

\subsection{The comparison of the results of $\pi^2$-resummations.}
We now consider the result of applying these different
procedures for
resummations of
kinematic $\pi^2$-effects in the perturbative coefficients
for $\rm\Gamma_{H\overline{b}b}$. We will follow the
results obtained in  Ref.\cite{Bakulev:2008hx} and
discuss the comparison of  the behavior of
various parameterizations
for this quantity, which were  defined in the previous section,
with the $\alpha_s^2$- and $\alpha_s^3$- truncated
$\rm\overline{MS}$-scheme expression of Eq.(2.1).

In the process of these discussions we use plots, similar
to those, presented in   Ref.\cite{Bakulev:2008hx}, but modified
at our request by changing the values   of
the QCD parameters to the ones presented in Table 1. 
The results of these additional studies
were kindly communicated \cite{Bakulev:2009}  to us in the form
of the figures presented below.

\begin{figure}[h]
{\includegraphics[width=53mm]{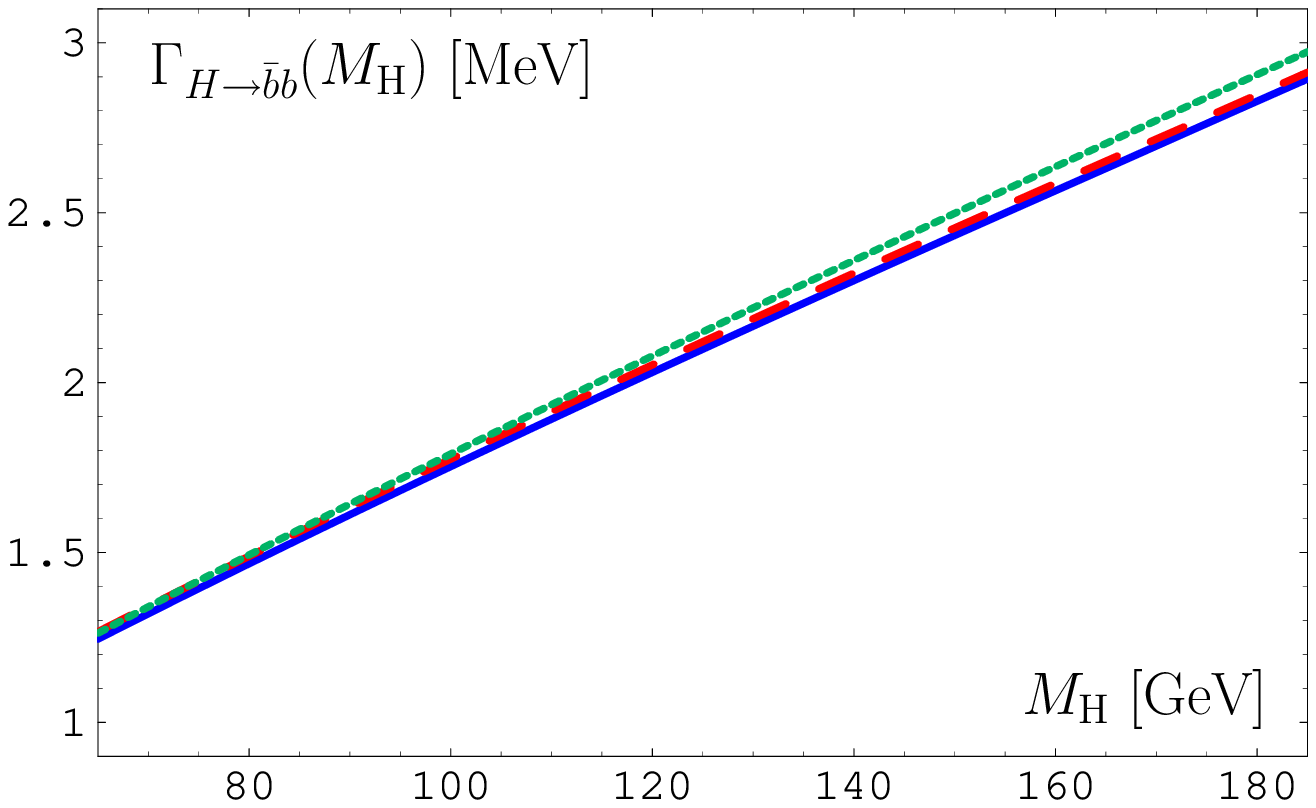}}
{\includegraphics[width=53mm]{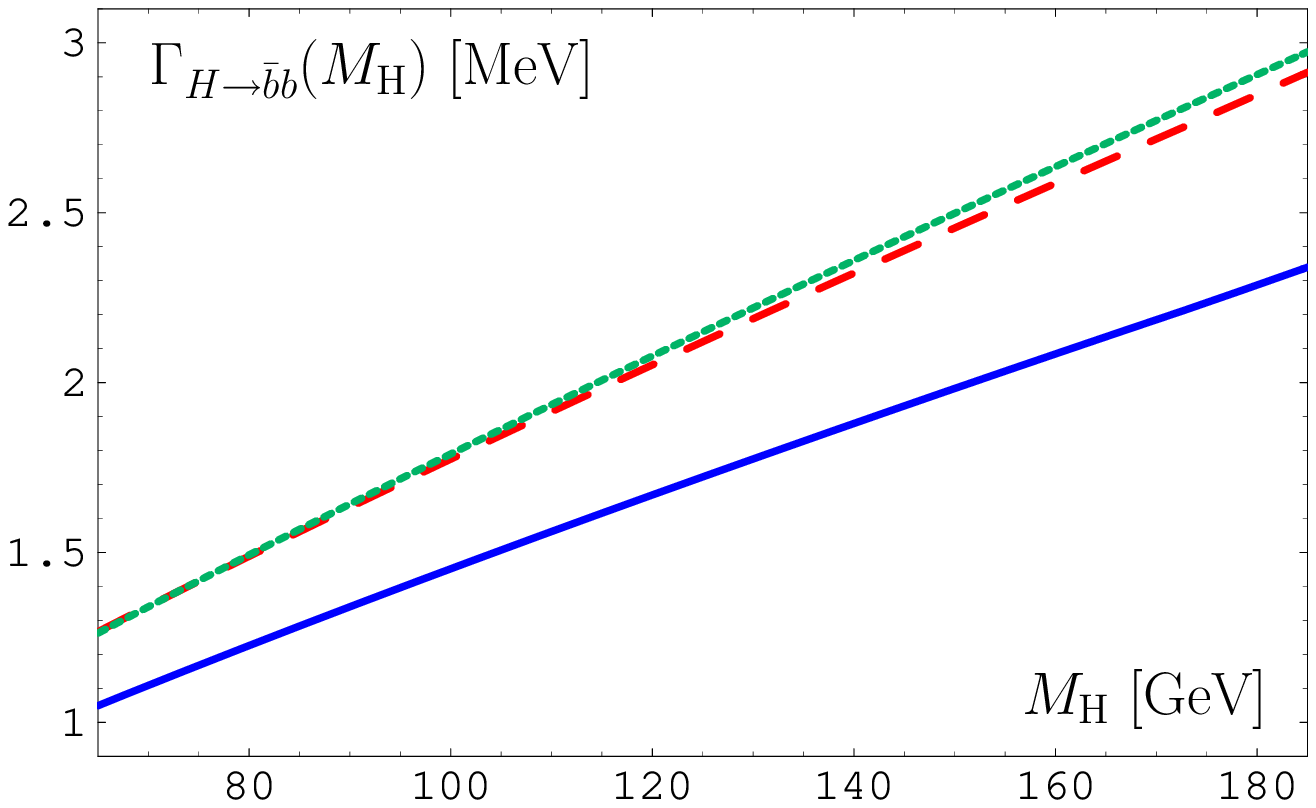}}
\caption{The comparison of N$^2$LO   approximations for
$\rm\Gamma_{H\overline{b}b}$.
On both panels the dashed line is the truncated
$\alpha_s^2$-approximation of the $\rm\overline{MS}$-scheme result.
The dotted line is the 1-loop FAPT expression
of Eq.(3.4), which is identical to the CIPT expression  of
Eq.(3.1).
The solid line on the left figure displays  the 2-loop FAPT
analog of Eq.(3.7), while on the right figure it corresponds to
2-loop variant of the ``flavour-corrected global  FAPT'' approximant of Eq.(3.8).}
\label{figB2}
{\includegraphics[width=53mm]{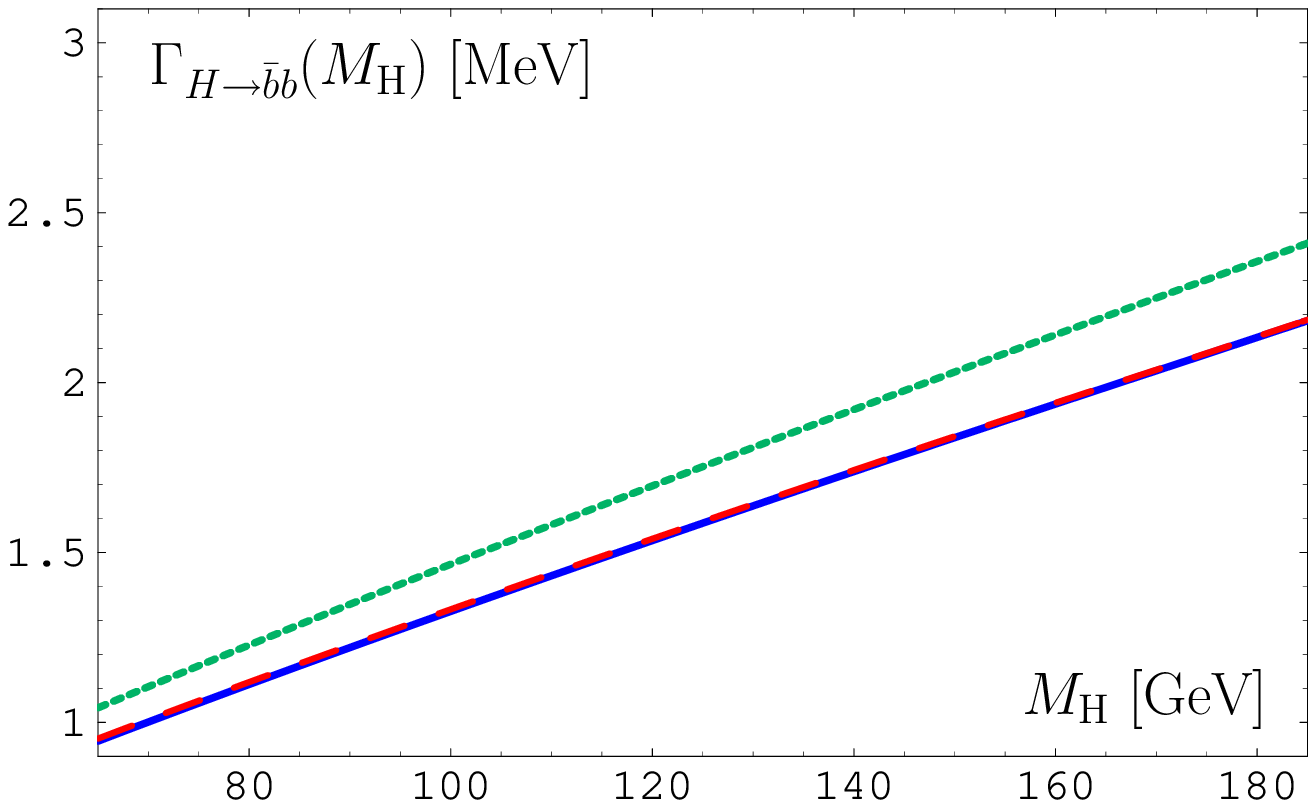}}
{\includegraphics[width=53mm]{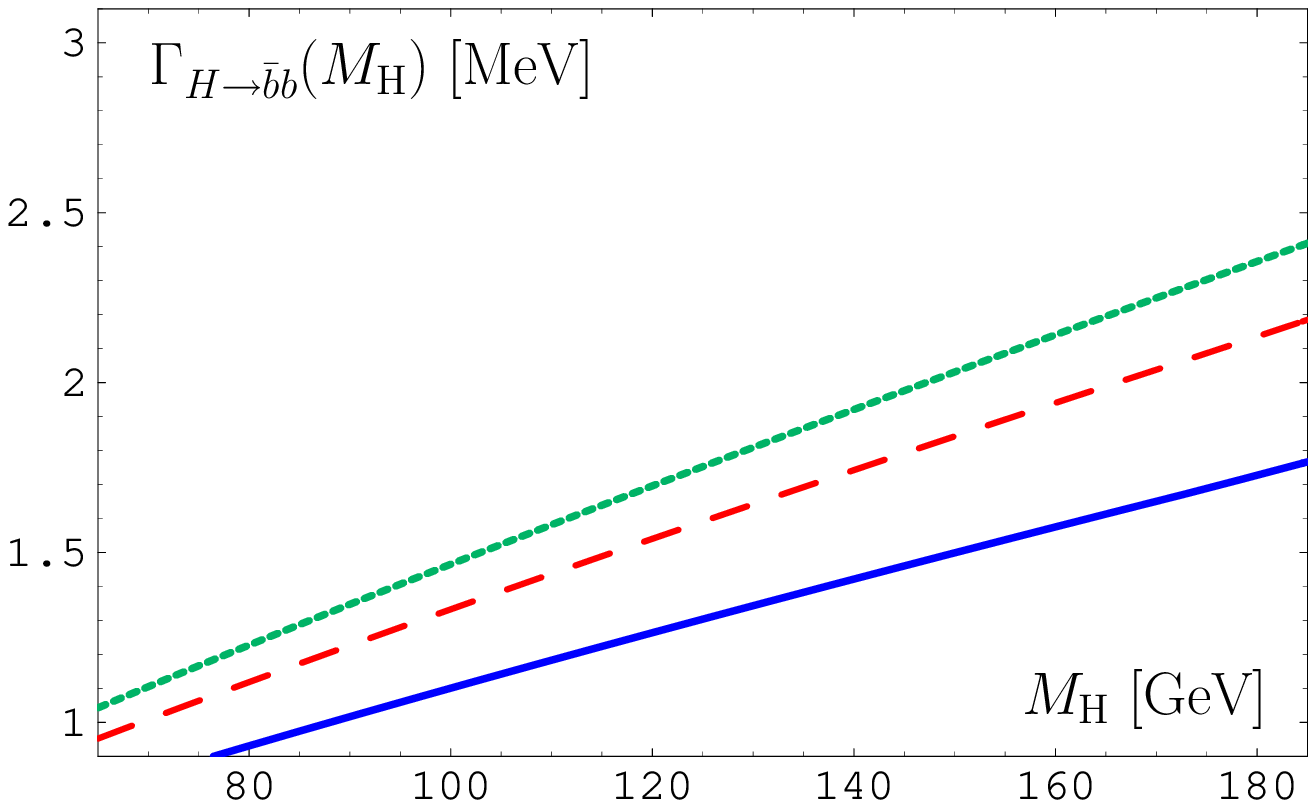}},
 \caption{The N$^3$LO version  of Fig.3. The dashed line is the truncated
$\alpha_s^3$ $\rm\overline{MS}$-scheme expression.}
\label{figB3}
\end{figure}

A careful look to the  curves from Fig.3 and Fig.4 leads us
to the following observations:
\begin{enumerate}
\item
at the N$^2$LO-level the resummation of the $\pi^2$-terms {\bf do not} lead
to detectable effects. The related curves almost coincide with the
$\alpha_s^2$ $\rm\overline{MS}$-scheme approximation  
for
$\rm\Gamma_{H\overline{b}b}$, obtained in Ref. \cite{Gorishnii:1990zu};
\item
at the N$^3$LO level the effects of resummation of the kinematic
$\pi^2$-contributions are visible. Indeed, for $\rm{M_H}=120~{\rm GeV}$
the value of the   resummed approximant is over 0.2 $\rm{MeV}$ lower, than
the value of the truncated $\alpha_s^3$ $\rm\overline{MS}$-scheme
expression   for $\rm\Gamma_{H\overline{b}b}$, obtained in
Ref.\cite{Chetyrkin:1996sr}. The behavior of the curves of the left
plot from  Fig.3 indicate that for a Higgs boson with this mass, 
the value of 
the  $\alpha_s^3$-approximation for  its decay width to $\overline{b}~b$
quarks  is  ${\rm\Gamma_{H\overline{b}b}\approx 1.7~MeV}$, while the
resummation of the $\pi^2$-effects decreases  this by over 11$\%$;
\item
an  obvious  message, which comes from the consideration of the right-hand
plots of Fig.3 and Fig.4 is that the application of   ``the 
flavor-corrected global  FAPT''  with full analytization 
of $\rm{n_f}$ dependence in $\rm\Gamma_{H\overline{b}b}$ leads to
smaller  values of the decay width for Higgs boson, 
than  in the case of the  truncated
$\rm\overline{MS}$-scheme approach, and  2-loop and
3-loop FAPT approach \cite{Bakulev:2006ex}.
An attempt to explain the phenomenological reason for this
reduction is now in progress \cite{Bakulev:2009}.
It may be also of interest to compare FAPT applications 
with the existing  applications to $\rm\Gamma_{H\overline{b}b}$ of 
the CIPT based   resummation procedure from Ref.~\cite{Maxwell:2001uv}.  
\end{enumerate}

{\bf Conclusions}.

In the discussions presented above we found that even for the fixed
values of the QCD parameters used in Table 1,  which do not take into
account the existing theoretical and experimental uncertainties
in the values of $\alpha_s(\rm{M_Z})$ (and thus
${\rm\Lambda_{\overline{MS}}^{(n_f=5)}}$), and of the  running
and pole $b$-quark masses, different parameterizations of the decay
width of the $H\rightarrow\overline{b}b$ process  deviate from the
$\rm\overline{MS}$-scheme prediction. In our view,  this
feature demonstrates the existence of additional theoretical
QCD uncertainties, which are not usually considered in 
phenomenological studies. We still do not know what is the
role of similar QCD uncertainties in the theoretical
predictions for other characteristics of  Higgs boson
production and its other   decays (see however, the recent
studies of resummation of terms 
proportional to  $\pi^2$ in the N$^2$LO  perturbative QCD  predictions
for the Higgs -boson production cross-section  and for the 
hadronic decay rate $\rm\Gamma(H\rightarrow gg)$ \cite{Ahrens:2008nc}).
It will also be interesting to understand whether these
uncertainties survive in the branching ratios of the  Standard Model Higgs
boson for different values of its mass.

{\bf Acknowledgments}

We are grateful to G.~Altarelli, A.~P.~Bakulev, M. Beneke, M.~Yu.~Kalmykov,  
L.~N. Lipatov,  C~.J~. Maxwell, S.V. Mikhailov,  
A.~A.~Penin and N.~G. Stefanis  for useful discussions and comments on 
various 
topics, related to this work. One of us (A.~L.~K.) wishes to thank the
members of the Organizing Committee
 of the  XII Workshop on  Advanced Computing and Analysis Technique
in Physics Research for the invitation to present the plenary talk
and for the hospitality in Erice, Sicily,  Italy. The write-up  on this talk
was started when the both of us were visiting CERN Theoretical Physics 
Unit, which we thank for their kind hospitality.

\end{document}